\documentclass{aa}
\usepackage{graphicx}
\def\gapprox{\lower.4ex\hbox{$\;\buildrel >\over{\scriptstyle\sim}\;$}}
\def\lapprox{\lower.4ex\hbox{$\;\buildrel <\over{\scriptstyle\sim}\;$}}
\begin{document}

%\thesaurus{02.01.1,02.18.5,11.01.2,11.10.1,13.18.1,13.25.2}

\title {Are the hotspots of radio galaxies the sites of {\it in-situ}
acceleration of relativistic particles?}

\author{Gopal-Krishna\inst{1} \and P.\ Subramanian\inst{1}\fnmsep\inst{2}
\and P.\ J.\ Wiita \inst{3}\fnmsep\inst{4} \and P.\ A.\ Becker\inst{5}}

\offprints{Gopal-Krishna}

\institute{National Centre for Radio Astrophysics/TIFR, 
Pune University
Campus, Post Bag No.\ 3, Ganeshkhind, Pune 411007, India 
({\tt krishna@ncra.tifr.res.in}) 
\and
Inter University Centre for Astronomy \& Astrophysics,
Pune University Campus, Post Bag No.\ 4, Ganeshkhind, Pune 411007, India
({\tt psubrama@iucaa.ernet.in})
\and  Department of Physics and
Astronomy, Georgia State University, University Plaza, Atlanta, GA 30303-3083, USA
 ({\tt wiita@chara.gsu.edu})
\and
Department of Astrophysical Sciences, Princeton University, Princeton NJ
08544-1001, USA 
\and
Center for Earth Observing and Space Research, George Mason University,
Fairfax, VA 22030, USA ({\tt pbecker@gmu.edu})}

\date{Received 5 February 2001; Accepted 15 August 2001}

\authorrunning{Gopal-Krishna et al.}
\titlerunning{Are hotspots acceleration sites?}
%\maketitle
\markboth{}{}

\abstract{
Using a large set of optically detected hotspots in powerful extragalactic 
double radio sources, we examine the basic question of whether the detection 
of optical synchrotron radiation requires {\it in-situ} acceleration of 
relativistic electrons within the hotspots/lobes. For this, we take into
account the jet's bulk relativistic motion, as well as its likely 
misalignment from the plane of the sky. Together, both these factors 
can drastically
 reduce the apparent range of the  ultra-relativistic electrons
ejected from the nucleus in the form of a jet. The
conventionally adopted parameter space for the fundamental variables, 
namely, the hotspot magnetic field, radio source orientation angle 
relative to the line-of-sight and the bulk speed of the jet plasma, is
considered. We find that the observed optical/near-IR synchrotron 
emission of the hotspots can be explained even if the radiating relativistic 
electrons were accelerated exclusively within the nuclear 
region, provided the energy losses incurred by the electrons during 
their transport down the jet are dominated by inverse Compton upscatterings 
of the cosmic microwave background photons. Under this 
circumstance, {\it in-situ} acceleration of relativistic electrons inside 
the hotspots or lobes is not found to be mandated by their reported 
optical/near-infrared detections. 
}

\maketitle

\section{Introduction}

A large fraction of the extra-nuclear radiation detected from radio 
galaxies is generally believed to arise from ultra-relativistic electrons
with Lorentz factors $\gg$ 100, undergoing synchrotron and/or inverse Compton
losses.  It is now well established that these electrons are initially
accelerated in the vicinity of a compact supermassive object located at 
the active galactic nucleus, from where they are transported out to 
kiloparsec, or even megaparsec, distances in the forms of collimated 
jet-like features, giving rise to a pair of lobes of nonthermal emission 
on the opposite sides of the nucleus (e.g., Begelman et al.\ \cite{bege84};
 see, however, Valtonen \& Hein\"am\"aki \cite{valt}, for an alternative 
scenario).
In powerful radio galaxies and quasars, the jets are seen to terminate 
in compact, bright regions, called {\it hotspots}, wherein much of the ordered 
kinetic power of the jet is presumed to be converted into random
motion within the relativistic 
plasma and strengthened magnetic fields. 

A long debated key issue is whether the ultra-relativistic 
electrons radiating within the kiloparsec-scale hotspots and lobes are 
largely directly supplied by the nucleus in the form of jets, or whether 
a significant amount of relativistic particle acceleration occurs 
{\it in situ} within the hotspots and lobes, for instance, near the 
shock fronts where the jet flow is thermalized (e.g., Scheuer \cite{sche89}; 
Blandford \& Rees \cite{blan}). To effectively probe this 
question, many deep searches for optical/near-infrared (and even X-ray) 
radiation from hot spots have been carried out, motivated by the fact 
that the required electron Lorentz factors $(\gamma > 10^5)$ correspond 
to a lifetime 
of $\lapprox 10^3$ yr against synchrotron losses, even in a magnetic field 
as weak as a few microgauss (e.g., Lelievre \& Wlerick \cite{leli}; 
Kronberg \cite{kron76}; Saslaw et al.\ \cite{sasl}; Simkin \cite{simk78};
Meisenheimer \& R\"oser \cite{meis86}; Crane et al.\ \cite{cran}; 
R\"oser \& Meisenheimer \cite{rose87}). 
While such early searches already led to positive detections of optical
counterparts of a few radio hotspots, the subsequent detection of 
linear polarization firmly established the synchrotron nature of the
optical hotspots (e.g., Meisenheimer \& R\"oser \cite{meis86}; 
R\"oser \& Meisenheimer \cite{rose87}; Hiltner et al.\ \cite{hilt}; 
Thomson et al.\ \cite{thom}; L\"ahteenm\"aki \& Valtaoja \cite{laht}). 
X-ray detections of a few hot spots with {\it ROSAT} and 
{\it CHANDRA} have also been reported. However, 
in the absence of linear polarization measurements, attributing
a synchrotron origin to the X-ray emission is premature; in fact, an 
inverse-Compton
origin or a synchrotron self-Compton explanation for the the X-ray emission 
is usually favored (e.g., Harris et al.\ \cite{harr94}, \cite{harr00}; Hardcastle et al.\
\cite{hard}; Wilson et al.\ \cite{wils00};
see, however, Section 3 for the case of Pictor A radio 
lobe detected recently with {\it CHANDRA}).

In a detailed analysis of the radio-to-optical synchrotron spectra of 8 
hot spots associated with powerful,  FR II   (Fanaroff \& Riley \cite{fana})
radio galaxies, Meisenheimer et al.\ (\cite{meis97})
 argued that the observed spectral cut-off in the 
frequency range $\nu_c = 3\times 10^{11} - 4\times 10^{14}$ Hz can often be 
satisfactorily explained in terms of diffusive shock acceleration of 
relativistic electrons (up to the required $\gamma \sim 10^5$) at the 
strong terminal shock coincident with the hotspot. The theoretical 
framework for the {\it in-situ} shock acceleration incorporating 
first-order Fermi acceleration has been developed in detail by many 
authors (e.g., Bell \cite{bell}; 
 Biermann \& Strittmatter \cite{bier}; Drury \cite{drur}; Heavens \& Meisenheimer \cite{heav};
Kirk \& Schneider \cite{kirk}). 
Some alternative proposals for production of ultra-relativistic
electrons/positrons within the radio jets and lobes involve tapping, 
by various physical processes, the energy contained in a possible relativistic 
proton component of the beam flow (Dar \& Laor \cite{dar}; 
Mannheim et al.\ \cite{mann91}; Mannheim \cite{mann93}; Mastichiadis \& Kirk \cite{mast}; 
Subramanian et al.\ \cite{subr}), or a putative neutron component (Eichler \& Wiita \cite{eich}; 
Contopoulos \& Kazanas \cite{cont}), or simply,
low-frequency electromagnetic waves (Rees \cite{rees71}).

On the other hand, any model in which ultra-relativistic electrons are
generated solely within the active nucleus must ensure their survival
against the radiative losses suffered during their transit to the hotspot.
Of the two principal energy loss mechanisms, namely, synchrotron and
inverse-Compton emission, even the theoretically {\it minimum loss}
scenario cannot evade the radiative losses arising from inverse-Compton
scattering of the cosmic microwave background (CMB) photons. In this 
communication, we critically examine this question by taking into account 
the role of bulk relativistic motion of the jets, which is now widely 
believed to be maintained all the way out to multi-kpc-scales (e.g., Scheuer \cite{sche87};
 Laing \cite{lain88}; Garrington et al.\ \cite{garr}; 
Bridle et al.\ \cite{brid94}; Bridle \cite{brid96}; Biretta et al.\ \cite{bire}) and 
which would lead to Doppler-enhanced CMB scattering losses.  The 
visualization of the jets as the conduits of energy supply to the 
hotspots, in a manner largely free from synchrotron losses, has been 
discussed by several authors (e.g., Shklovskii \cite{shkl}; Kundt \cite{kund86};
Owen et al.\ \cite{owen}; Eilek et al.\ \cite{eile}), although its viability  has been 
questioned by some 
others (e.g., Felton \cite{felt}; Meisenheimer \cite{meis96}).

Nonetheless, since our aim here is to check if the
need for {\it in-situ} acceleration of electrons inside the hot spots
is inescapable,
we shall  confine our attention to the energy losses that are 
absolutely unavoidable: the losses due to scattering of the CMB photons 
by the relativistic electrons flowing in the jet. (Note, however, 
that a minor loss of the jet power {\it via} synchrotron emission,
particularly through interactions at the surface of the jet,
 is fully consistent with this
picture; Sect.\ 3). For visualizing the energy transport we shall confine
ourselves to the conventional picture, whereby a jet consists of a
quasi-continuous train of synchrotron plasmons in bulk
relativistic motion, which radiate isotropically in their own frame
of reference. It is also widely believed 
that the speeds with which the hotspots 
advance are non-relativistic (e.g., Scheuer \cite{sche95}; Arshakian \& Longair \cite{arsh})
and, consequently, any Doppler correction 
to their observed radiation is negligible. 
This avoids some of the uncertainties encountered while
interpreting the jet emission.

We now focus attention on the thirteen radio galaxies and quasars for 
which the published high-resolution images have allowed determination of 
the radio--optical spectrum for at least one of the hotspots, thereby 
yielding a useful estimate of the spectral break frequency, $\nu_c$ 
(Table 1).  When optical and/or IR fluxes are found to be well registered with
radio images of the hotspots, $\nu_c$ can be fit to the radio
through optical spectra
using the method described in 
Meisenheimer et al.\ (\cite{meis89}).  If the photometric measurements 
are made at multiple frequencies
bracketing $\nu_c$, its value can be determined to better than 50\%
 (what 
we call Class I measurements), although
if the measurements are further from $\nu_c$, the error in its 
determination could be up to a factor of 2 to 3 (Class II); since our
results for the maximum distances to which plasma can propagate
only depends weakly on $\nu_c$ (Eq.\ 6), such uncertainties are
acceptable.
In a few cases, where multi-frequency photometric data in the 
crucial optical/near-IR region are not available, we have taken, as a 
lower limit to $\nu_c$, $50\%$ of the highest (optical/near-IR) frequency 
at which the hotspot has been detected; (these comprise the
least certain, Class III, measurements).  Still
this assumption seems reasonable 
considering the steep decline that is seen to set in at ${\nu_c}$ in
the spectra of well detected hotspots (e.g., Meisenheimer et al.\ \cite{meis97}). 
Our sample consists of 15 hotspots which are the prominent 
cases of optical synchrotron detections reported in the literature; 
however, a complete coverage of this literature is not claimed.
In column (1) of Table 1 we give the source names, in column (2) whether it
is a quasar (Q) or a radio galaxy (G), along with its redshift, $z$. Columns (3),
(4) and (5) give the equivalent magnetic field of the cosmic microwave background,
measured in the frame of the jet for the three adopted values of the 
jet bulk Lorentz factor. Column (6) provides the angular separation of the 
hotspot from the nucleus. Columns (7)
and (8) give the corresponding linear separation for different cosmological models.
Column (9) notes whether polarization was detected in the optical hotspot, thereby
cementing its synchrotron origin.  Column (10) lists our best value for $\nu_c$,
and column (11) our estimate of the reliability of that value.

\section {The model}

A relativistic electron with Lorentz factor $\gamma$
 on its way from the galactic nucleus to the hotspot loses energy
to inverse Compton collisions with CMB photons of energy
density, $u_{\rm rad,j}$, and also suffers synchrotron losses in the 
transverse component of the
jet field, $B_{j}$.  The energy-loss timescale of such an electron is 
 (Longair \cite{long}; Meisenheimer et al.\ \cite{meis89})
\begin{equation}
\tau_{\rm j} = 2.01 \times 10^{11}  \gamma^{-1} 
\biggl[\frac{0.33  u_{\rm
rad,j}}{10^6 {\rm eV}~{\rm m}^{-3}} + 
\biggl (\frac{B_{j}}{\rm nT} \biggr )^{2} \biggr]^{-1} {\rm yr}.
\end{equation}
We stress that the quantities $u_{\rm rad,j}$, $\tau_{\rm j}$, and $\gamma$
are  measured in the jet frame.  The energy density of the CMB  photons in the
`laboratory' frame (the frame in which the CMB radiation is isotropic) is
(Scott et al.\ \cite{scot}):
\begin{equation}
u_{\rm rad} = 2.62 \times 10^5 \, (1 + z)^{4} \, \, {\rm eV} \, {\rm m}^{-3} \, ,
\end{equation}
where $z$ denotes the cosmological redshift. In the frame of the jet, which is
moving with bulk Lorentz factor $\Gamma_j$, the energy density of the
impinging  CMB 
photons is (e.g., Dermer \& Schlickeiser \cite{derm93}, \cite{derm94}; Begelman et al.\
\cite{bege94}):
\begin{equation}
u_{\rm rad,j} \simeq \Gamma_{j}^{2} \, u_{\rm rad} \, .
\end{equation}
The equivalent  magnetic field corresponding to $u_{\rm rad,j}$ 
is $B_{\rm CMB}$(nT) = 
$0.325\, \Gamma_{j} \, (1+z)^{2}$ and is given in Table 1
for $\Gamma_j = 2$, 5 and 10.

Now, the synchrotron loss timescale, $\tau$, in the `laboratory' frame is 
related to that in the jet frame by the Doppler factor $\delta_j$:
\begin{equation}
\tau = \delta_j^{-1}\,\tau_{\rm j} \equiv \Gamma_{j}\,(1 -
\beta_{j}\,{\rm cos}\,{\theta})\,\tau_{\rm j} \, ,
\end{equation}
where $\beta_{j} = (1 - \Gamma_{j}^{-2})^{1/2}$ and $\theta$ is the angle 
between the line-of-sight and the direction connecting the core and the hotspot.
In order to estimate the maximum range of the relativistic electrons, we follow
Meisenheimer et al.\  (\cite{meis89}) in expressing the electron 
Lorentz factor, $\gamma$, in
terms of ``observables'',  using $\nu_{c} = 42\,(B_{\rm HS}/{\rm nT})\,
\gamma^{2}$, where $\nu_{c}$ is the  break frequency in the frame of
the hotspot and $B_{\rm HS}$ is the magnetic field in the hotspot. 
(An averaging over pitch angles would yield a coefficient of
33 in place of 42 and would reduce the following
results for $D_{\rm max}$ by 11\%.)
If we now set $B_{j} = 0$ (i.e., neglect synchrotron losses along
the jet; cf. Sect.\ 1; see, however, Sect. 3), the maximum apparent 
(transverse) distance the jet's electrons can cover 
  before dropping to less than $e^{-1}$ of their initial
energy is given by:
\begin{equation}
D_{\rm max} = \tau v_{\rm app} =  \tau \frac{c\,\beta_{j}\,{\rm
sin}\,\theta}{(1 - \beta_{j}\,{\rm cos}\,{\theta})}.
\end{equation} 
We have used the usual expression for apparent transverse speed derived
for explaining the superluminal expansion of VLBI cores (e.g., Rees \cite{rees67}).

 Substituting Eqs.\ (1--4) in Eq.\ (5) and solving for $\gamma$
in terms of $\nu_c$ and $B_{\rm HS}$, we get
\begin{equation}
D_{\rm max} = 145\, \frac{\beta_{j} \, {\rm sin}\,\theta } {\Gamma_{j} \, 
(1+z)^{4.5}}\,\biggl [ \frac{(B_{\rm HS}/{\rm nT})}{(\nu_{c}/10^{15}\,{\rm Hz})} 
\biggr ]^{1/2} \, \, {\rm kpc}.
\end{equation}
Note that in Eq.\ (6), we have included in the denominator a factor of
$(1+z)^{1/2}$ to correct for the cosmological redshift of
$\nu_{c}^{1/2}$.  More precise calculations of Eq.\ (3) and the
rate of energy loss, given in the Appendix,
show that the results in Eqs.~(4--6) should be multiplied by a factor 
of\break
$(1+\beta_j^2/3)/(1+\beta_j^2)$, which is always between $2/3$ and
$1$.

To recapitulate, in the present analysis we have 
improved on the earlier treatments by making an allowance for both
(i) the increased CMB energy density experienced by the jet plasma
due to its bulk relativistic motion, and (ii) the offset of the radio 
axis from the plane of the sky, which is non-negligible in general.
Consistent with the unified scheme for FR II radio sources (e.g., Barthel \cite{bart};
 Gopal-Krishna et al.\ \cite{gkkw}), the approximate range of $\theta$ is $45^{\circ} - 90^{\circ}$ for radio 
galaxies and $20^{\circ} - 45^{\circ}$ for quasars, with blazars
corresponding to the smallest angles.

\begin{figure*}
\centering
\includegraphics[width=17cm]{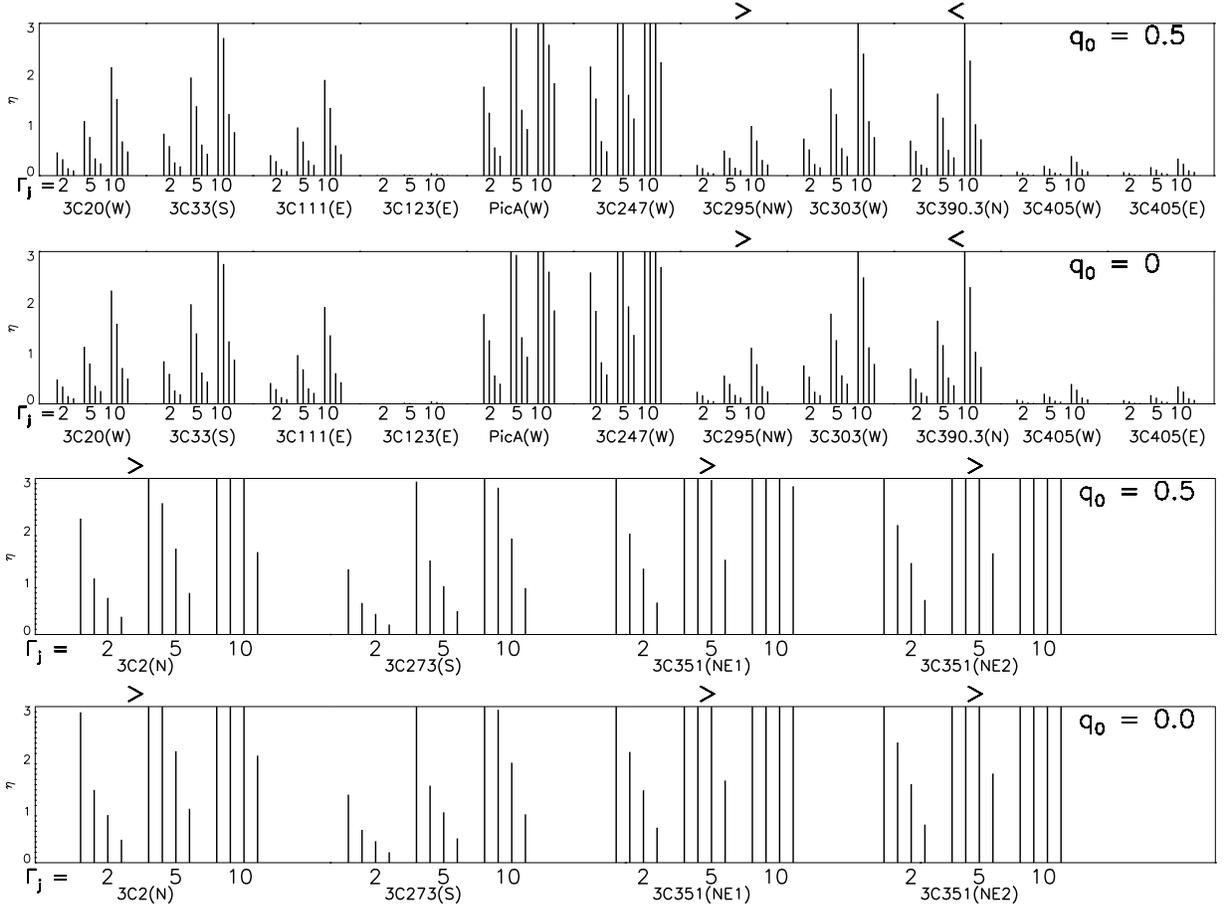}
%\resizebox{\hsize}{!}{\includegraphics{fig1.ps}}
\caption{The computed 
values of $\eta = D_{\rm obs}/D_{\rm max}$  including the relativistic
correction derived in the Appendix, for all 15 hotspots in our sample.
The top two panels show the results for radio galaxies (for two
values of $q_0$), while the results for the quasars are displayed in the
bottom two panels. For each hotspot, there are three sets of four vertical
lines each, representing the three values of the jet's bulk Lorentz 
factor $\Gamma_{j}$. In the top two panels, each set of four vertical
lines has the following combinations of $\theta$ and $B_{\rm HS}$, from 
left to right: ($\theta = 45^{\circ}$, $B_{\rm HS} = 10$ nT),
($\theta = 90^{\circ}$, $B_{\rm HS}  = 10$ nT), 
($\theta = 45^{\circ}$, $B_{\rm HS} = 100$ nT), 
($\theta = 90^{\circ}$, $B_{\rm HS} = 100$ nT). 
The corresponding values for the bottom two panels are: 
($\theta = 20^{\circ}$, $B_{\rm HS} = 10$ nT),
($\theta = 45^{\circ}$, $B_{\rm HS} = 10$ nT), 
($\theta = 20^{\circ}$, $B_{\rm HS} = 100$ nT),
($\theta = 45^{\circ}$, $B_{\rm HS} = 100$ nT) (see Sect. 2).  
For cases where only a lower limit to $\nu_c$ is available (Table 1),
the value of $\eta$ is a lower limit, as indicated by the symbol $>$
near the top, and vice versa. Note that all values of $\eta > 3$ 
are truncated at $\eta = 3$ in the plots.
}
\label{fig1}
\end{figure*}

In computing $D_{\rm obs}$ in Table 1, a Hubble constant 
$H_0$ = 75 km s$^{-1}$ Mpc$^{-1}$ 
is assumed and two values of $q_{0}$ (= 0.5 and 0) are considered. We have used the
usual relation (e.g., Lang \cite{lang})
\begin{equation}
D_{\rm obs} = \frac{c \, \Theta}{H_0} 
\frac{ \left\{q_0 z + (q_0-1)[(1+2 q_0 z)^{1/2}-1] \right\} } 
{q_0^2 (1+z)^2} \, ,
\end{equation}
to relate $\Theta$, the observed angular separation between the central object and the
hotspot (Col.\ 4 of Table 1), to $D_{\rm obs}$ (Col.\ 5 of Table 1), the linear separation 
between the two, measured at the same redshift as is $D_{\rm max}$. 

Using the data given in Table 1, we now compute for each hotspot the 
ratio $\eta \equiv D_{\rm obs}/D_{\rm max}$, taking the usually adopted limiting 
values of the parameters $\theta$, $B_{\rm HS}$ and $\Gamma_j$.
The limiting
values adopted for $B_{\rm HS}$ are 10 nT and 100 nT, in that 35 nT is the
typical magnetic field strength in the hotspots (e.g., 
Meisenheimer et al.\ \cite{meis97}).
For $\Gamma_j$ we consider three values:  $\Gamma_j$ = 2, 5 and 10, which
cover the range adopted in conventional models of the kpc-scale jets
(e.g., Orr \& Browne \cite{orr}; Vermeulen \& Cohen \cite{verm}; Bicknell \cite{bick};  
Wardle \& Aaron \cite{ward}).  The computed values of $\eta$ for different
combinations of the input parameters are plotted for each hotspot in
Fig.\ 1.   The relatively weak dependence of
$D_{\rm max}$ on $\nu_c$ means that even rather large errors
in determining the break frequency do not have a major effect
on our results.

\cleardoublepage
{\begin{table}
\caption[]{Properties of the optical/near-infrared hotspots in the 
FRII radio sources$^{*}$}
\label{sphericcase}
\begin{tabular}{ccccccccccc}
\hline
Source & \multicolumn{1}{c} {Opt. ID} & \multicolumn{3}{c}{$B_{\rm
CMB}$(nT)} &
\multicolumn{1}{c}{$\Theta$($^{\prime \prime}$)} &
\multicolumn{1}{c}{D$_{\rm obs}$(kpc)} &\multicolumn{1}{c}{D$_{\rm
obs}$(kpc)} & \multicolumn{1}{c}{Opt. Pol.} &
\multicolumn{1}{c}{$\nu_{c}$(Hz)} & \multicolumn{1}{c}{Class$^{\dag}$}\\
Hotspot &z(ref) & $\Gamma_{j}=2$ & $\Gamma_{j}=5$ & $\Gamma_{j}=10$ &
(ref) &$q_{0}=0.5$&$q_{0}=0$& (ref) & (ref) & \\
(1)&(2)&(3)&(4)&(5)&(6)&(7)&(8)&(9)&(10)&(11)\\
\hline
\\
0003-003 &Q&2.7&6.7& 11.0 &$0.83$ & 4.7 & 6.1 & -- & $\gapprox 9 \times 10^{14}$ & II\\
3C2 (N) &1.037(a)& &  & &(a,b)& & & & (a) & \\
 & & & & &  \\
0040+517 &G&0.9&2.2&3.7& $24.7$& 62.9 & 65.7 & detected & $\sim 1.3 \times 10^{14}$ & I\\
3C20 (W) &0.174(c)& & & &(c)& &  & (d) & (c) & \\
 & & & & &  \\
0106+130 &G&0.7&1.8&3.0&$114$& 118 & 120 & detected & $\sim 3.0 \times 10^{14}$ & I\\
3C33 (S) &0.0592(e)& & & &(c,f,g)& & & (g) & (c) & \\
 & & & & &  \\
0415+379 &G&0.7&1.8&2.9&$122$& 105 & 107 & detected & $\sim 1 \times 10^{14}$ & I\\
3C111 (E) &0.0485(c)& & & &(c,h,i)& & & (j) & (c) & \\
 & & & & &  \\
0433+295 &G&1.0&2.4&3.9&$8.6$& 25.7 & 27.2 & -- & $\sim 3\times 10^{11}$ & I \\
3C123 (E)$^{1}$ &0.2177(c)& & & &(c,h,k,l)& & & & (c) & \\
 & & & & &  \\
0518-458 &G&0.7&1.7&2.8&$252$& 161 & 163 & detected & $\sim 9 \times 10^{14}$ & I\\
Pic A (W)$^{2}$ &0.0350(c)& & & &(m,c,h,n)& & & (o) & (c) & \\
 & & & & &  \\
1056+432 &G&2.0&5.0&8.1&$\sim 9$& 49 & 59 & -- & $\sim 1.3 \times 10^{14}$ & II \\
3C247 (W) &0.749(p)& & & &(q,r)& & & (q) & (q) & \\
 & & & & &  \\
1226+023 &Q&0.9&2.2&3.6&$21.3$& 50.4 & 52.5 & detected & $\sim 4 \times 10^{14}$ & I \\
3C273 (S)$^{3}$ &0.158(c)& & & &(s)& & & (t) & (c,h,s) & \\
 & & & & &  \\
1409+524 &G&1.4&3.5&5.7&$1.9$& 8.7 & 9.8 & -- & $\gapprox 2 \times 10^{14}$ & III$^{\ddag}$\\
3C295 (NW)$^{4}$ &0.461(u)& & & &(u)& &  & (u) & (u) & \\
 & & & & &  \\
1441+522 &G&0.9&2.2&3.5&$16.6$& 36.0 & 37.3 & $3 \sigma$ & $\sim 1.3 \times 10^{15}$ & II\\
3C303 (W)$^{5}$ &0.141(c)& & & &(c,v,q,w)& &  & (j) & (c) & \\
 & & & & &  \\
1704+608 &Q&1.2&3.1&5.0& & & & & &\\
3C351 (NE1)$^{6}$ &0.371(j)& & & &$25.8$&106 & 117 & $2 \sigma$ & $\gapprox 2 \times 10^{14}$ &
III$^{\ddag}$ \\
3C351 (NE2)$^{6}$ &0.371(j)& & & &$28$& 116 & 127 & $2 \sigma$ & $\gapprox 2 \times 10^{14}$ &
III$^{\ddag}$ \\
 & & & &  &(j,w) & & & (j) & (j) & \\
 & & & & & \\
1845+797 &G&0.7&1.8&3.0&$101$& 99.7 & 101 & $1 \sigma$ & $\lapprox 3 \times 10^{14}$ & II \\
3C390.3 (N)$^{7}$ &0.0570(x)& & & &(x,y)& & & (j) & (y,z,j) & \\
 & & & & & \\
1957+406 &G&0.7&1.8&3.0& & & & &\\
3C405 (W)$^{8}$ &0.0565(c)& & & &$67$& 66.1 & 67.0 & -- & $\sim 10^{13}$ & II \\
Cyg A (E)$^{8}$ &0.0565(c)& & & &$58$& 57.2 & 58.0 & -- &$\sim 10^{13}$ & II \\
 & & & & &($\alpha$) & & & & (c,$\beta$,$\gamma$) & \\
\hline
\end{tabular}
\end{table}

\clearpage
%\cleardoublepage
\noindent {\small{\bf Notes to Table 1:}}

\noindent \small{*: We have assumed a Hubble constant
$H_0$ = 75 km s$^{-1}$ Mpc$^{-1}$.}

\noindent \small{\dag: 
Classes I, II and III represent decreasing confidence in the
estimate of $\nu_{c}$, depending upon the available spectral coverage
for the hotspot.
Class I denotes the cases where the 
measurements bracket $\nu_{c}$ fairly tightly; 
class II where the multiple measurements lie to one 
side of $\nu_{c}$, whereas for 
class III, the hotspot is detected only at a single optical/near-IR
 frequency (Sect.\ 1).} 

\noindent \small{\ddag: The adopted value of $\nu_{c}$ is half the 
highest (optical/near-IR) frequency at
which the hotspot has been detected (Sect.\ 1)}\\ 
\small{1: Detected only up to 230 GHz (Meisenheimer et al.\ \cite{meis97})}\\
\small{2: Detected up to $\sim$ 5 keV (Wilson et al.\ \cite{wils01})}\\
\small{3: Detected up to $\sim$ 5 keV (Harris \cite{harr01}; R\"oser et al.\ \cite{rose00})} \\
\small{4: Detected up to $\sim$ 5 keV (Harris et al.\ \cite{harr00})}\\
\small{5: Resolved in I and K-band images (L\"ahteenm\"aki \& Valtaoja, \cite{laht};
 Meisenheimer et al.\ \cite{meis97}; Keel \cite{keel88}); there
may be a knot in the jet (Meisenheimer et al.\ \cite{meis97}).} \\
\small{6: Resolved in the I-band image; a double 
hotspot (L\"ahteenm\"aki \& Valtaoja, \cite{laht})}.\\ 
\small{7:
Resolved in I-band
image ($\sim 2^{\prime \prime}$, 
L\"ahteenm\"aki \& Valtaoja, \cite{laht}; Saslaw et al.\ \cite{sasl}). 
The spectrum is found to extend up to $\sim$ 2 keV (Prieto \& Kotilainen \cite{prie}; 
Harris et al.\ \cite{harr98}). 
However, a spectral 
break is apparent near $3 \times 10^{14}$Hz
(Harris et al.\ \cite{harr98}; Prieto \& Kotilainen \cite{prie}). 
This optical/near-IR hotspot appears to result from a collision of 
the jet with a dwarf galaxy which
deflects the jet (Leahy \& Perley \cite{leah}; Harris et al.\ \cite{harr98}).}\\
\small{8: Neither the eastern nor the western hot 
spot has a clear optical detection or a K-band
counterpart (Meisenheimer et al.\ \cite{meis97}). 
Both hotspots are detected in soft X-rays up to 5 keV 
(Wilson et al.\ \cite{wils00}; Harris et al.\ \cite{harr94}).}}

\noindent{\small {\bf References to Table 1:}
(a) Ridgway \& Stockton \cite{ridg};
(b) Saikia et al.\ \cite{saik};
(c) Meisenheimer et al.\ \cite{meis97};
(d) Hiltner et al.\ \cite{hilt};
(e) Simkin \cite{simk79};
(f) Dreher \& Simkin \cite{dreh};
(g) Meisenheimer \& R\"oser \cite{meis86};
(h) Meisenheimer et al.\ \cite{meis89};
(i) Linfield \& Perley \cite{linf};
(j) L\"{a}hteenm\"{a}ki \& Valtaoja \cite{laht};
(k) Okayasu et al.\ \cite{okay};
(l) Looney \& Hardcastle \cite{loon};
(m) Wilson et al.\ \cite{wils01};
(n) Perley et al.\ \cite{perl97};
(o) Thomson et al.\ \cite{thom};
(p) Polatidis et al.\ \cite{pola};
(q) Keel \& Martini \cite{keel95};
(r) Jenkins et al.\ \cite{jenk};
(s) R\"oser et al.\ \cite{rose00};
(t) R\"oser \& Meisenheimer \cite{rose91};
(u) Harris et al.\ \cite{harr00};
(v) Kronberg et al.\ \cite{kron77};
(w) Bridle et al.\ \cite{brid94};
(x) Leahy \& Perley \cite{leah};
(y) Prieto \& Kotilainen \cite{prie};
(z) Harris et al.\ \cite{harr98};
($\alpha$) Hargrave \&  Ryle \cite{harg};
($\beta$) Wilson et al.\ \cite{wils00};
($\gamma$) Harris et al.\ \cite{harr94}.}
%\cleardoublepage
%\end{table}

\section {Discussion}

In this study, our aim was to enquire if the reported observations of
optical/near-IR synchrotron emission from the hotspots of double radio sources
can be explained without having to invoke {\it in-situ} acceleration of 
relativistic electrons inside the hotspots. {\it In-situ} acceleration 
would be necessary if the parameter $\eta$ for a given hotspot were found 
to be well above unity for all combinations of the adopted plausible 
values of $\theta$, $B_{\rm HS}$ and $\Gamma_j$. It is seen from Fig.\ 1 
that for all the hotspots, {\it at least} one combination of the adopted 
plausible values of $\theta$ and $B_{\rm HS}$ yields $\eta \lapprox 1$, 
for values of $\Gamma_j$ up to 5. Even  
$\Gamma_j = 10$ is consistent with $\eta \lapprox 1$, 
with the possible exception
of the hotspots in 3C351; however, in both these cases the estimate of
$\nu_{c}$ is very uncertain (Class III, Table 1).
It is thus apparent that if the overall synchrotron losses of electrons
during their transit through the jet remain small compared with the
(inevitable) energy losses due to Compton upscatterings of the CMB photons,
the particles accelerated within the central nucleus can account for 
the synchrotron optical/near-IR emission detected from the hotspots.

The negligible amount of synchrotron losses within the jet, assumed
here, is in accord with some energy transport scenarios proposed earlier 
(e.g., Owen et al.\ \cite{owen}; Begelman et al.\ \cite{bege94}; Kundt \cite{kund96}; 
Heinz \& Begelman \cite{hein}). 
It is also consistent with the recent evidence in support of the large-scale
jets comprising a fast relativistic spine surrounded by a more dissipative,
slower moving layer of synchrotron plasma (e.g.,  Laing et al.\ \cite{lain99}; 
Swain et al.\ \cite{swai}; Bridle \cite{brid96}; also,  Chiaberge et al.\
\cite{chiab}; Giovannini et al.\ \cite{giov}; 
Carilli et al.\ \cite{cari};
Laing \cite{lain93}; Komissarov \cite{komi}),  although it should be noted that 
most of the evidence for this spine/sheath structure 
has been adduced for the weaker FR I sources.  Nonetheless, some FR II
sources do appear to behave similarly  (e.g.\ Chiaberge et al.\ \cite{chiaa}; Ghisellini
\& Celotti \cite{ghis}).

It may be emphasized 
that the main conclusion of this work does not demand assuming $B_j = 0$. 
 From Table 1, the equivalent magnetic field of the CMB, as would be
experienced by the relativistic jets of the sources in our sample,
has a mean value of $2.8$ nT. 
Thus, as long as the volume-averaged magnetic field is only up to 
$\sim 1$ nT,
the synchrotron losses in the jet will be minor in comparison to the
inverse Compton losses against the CMB photons. 
Here it may be recalled that
for the kpc-scale radio jets in moderately powerful radio galaxies, 
the typically adopted average magnetic field is $\sim 1$ nT 
(Bicknell \cite{bick}). Moreover, this value  
of $B_j$ is an average over the jet's cross section. If the magnetic
field is concentrated near the jet's outer layers due to surface interactions,
its strength there can be considerably
higher than 1 nT without forcing the synchrotron losses to become
the primary mechanism of energy loss in the overall jet flow.
Further, our picture invoking the dominance of inverse Compton losses 
within the jet obviously does admit dissipation within the jet. This is evident,
for instance, from the gradual spectral softening 
 along the 3C 273 
jet, as inferred from a comparison of its radio, optical and X-ray images 
(Harris \cite{harr01}).

By far the strongest case for {\it in-situ} acceleration reported in
the literature is that of the western lobe of the radio galaxy
Pictor A (Table 1). In this lobe, the linearly polarized optical
synchrotron radiation is seen to extend up to $\sim 10$ kpc from the
hotspot (Perley et al.\ \cite{perl97}; also, 
Thomson et al. 1995; R\"oser \& Meisenheimer 1987) and broadly co-spatial X-ray
 emission 
(of poorly understood origin)
has recently been detected by Wilson et al.\ (\cite{wils01}) using {\it CHANDRA}.
As noted by these authors, the level of X-ray emission is much above a 
simple extrapolation of the radio-optical spectrum, which shows a clear 
break in the near-IR region. Hence the X-ray flux cannot be  synchrotron
radiation from the same electron population. At the same time, 
inverse-Compton 
upscattering of the CMB photons by this electron population is also an 
unlikely explanation for the X-ray emission, for the X-ray spectral slope is 
found to be significantly different from that of the radio spectrum of the hotspot
(Wilson et al.\ \cite{wils01}). Thus, the source of the X-ray emission is
 unclear  at present. A similar situation may be present in the northern 
hotspot of 3C 390.3, whose V, R, and I photometry indicates a spectral cut-off 
near $3 \times 10^{14}$ Hz (Prieto \& Kotilainen \cite{prie}).  Hence any 
smooth extrapolation of this hotspots' spectrum would fall much 
below the level of  X-ray detection using $\it ROSAT$ (Harris et al.\ \cite{harr98}), 
which
is way above the level expected from Compton up-scatterings of either 
the internally generated photons or the CMB photons (Harris et al.\ \cite{harr98}; 
Prieto \& Kotilainen 1997).

At the same time, the large spatial extent of the optical synchrotron 
emission (up to $\sim 10$ kpc for Pictor A, see above) 
observed from a few radio lobes seems highly intriguing, given that the 
estimated synchrotron lifetimes of the optically radiating electrons in 
the lobe are only of the order of $10^3$ yr (R\"oser \& Meisenheimer \cite{rose87}; 
Perley et al. 1997), 
during which the electrons cannot travel 
more than $\sim 1$ kpc from the hotspot.  This apparent discrepancy may be 
resolved, if in reality, the magnetic field inside the lobes 
is concentrated in filaments, leaving large volumes with very weak field.
As emphasized by Scheuer (\cite{sche89}), relativistic electrons could then travel 
much farther away through the essentially field-free regions of the lobe, 
without decaying appreciably in energy, and thus retaining enough energy 
to generate widespread optical synchrotron emission upon interacting with 
magnetized filaments (see also, Gopal-Krishna \cite{gk}; Siah \& Wiita \cite{siah};
Tribble \cite{trib}; Eilek et al.\ \cite{eile}). This picture of 
highly filamentary magnetic field within the lobe is, in fact, supported 
by the radio fine structure detected in the VLA images of several radio 
galaxies. For instance, in the case of Cygnus A, only $\sim 10$\% of 
the lobe volume is estimated to be occupied by the magnetized filaments
(Perley et al.\ \cite{perl84}).

In summary, by adopting the generally accepted range of values 
for physical parameters of classical double radio sources 
(such as the 
hotspot's magnetic field, the jet's bulk speed and the radio axis 
orientation angle), we have argued that while {\it 
in situ} acceleration of relativistic electrons inside the 
hotspots/lobes of radio galaxies may seem an attractive possibility, 
the observations 
reported thus far do not make it a compelling requirement.

\begin{acknowledgements}
It is a pleasure to thank Rajaram Nityananda 
for fruitful discussions and the anonymous referee for helpful suggestions.
This work was supported in part by Research Program Enhancement
 funds at Georgia State University and by the
Council on Science and Technology at Princeton.  
\end{acknowledgements}

\appendix
\section{}
{A quantity of central importance in this paper is the timescale
$\tau$ for the ultrarelativistic electrons in the jet to cool via
the emission of inverse-Compton radiation. 
In the frame of the central object, which is the
`laboratory' frame, the jet has speed $c \beta_j$ and bulk
Lorentz factor $\Gamma_j = (1-\beta_j^2)^{-1/2}$, and the CMBR is isotropic. We wish to
calculate $\tau$ in this frame. The total inverse-Compton power
emitted per unit volume by an electron with Lorentz factor $\gamma$
and speed $c \beta = c \sqrt{1-\gamma^{-2}}$ as measured in the lab frame
is (e.g., Rybicki \& Lightman \cite{rybi})
\begin{equation}
P_{\rm compt}(\gamma) =\frac{4}{3} \, \sigma_T \, c \, \gamma^2
\, \beta^2 \, u_{\rm rad} \ ,
\end{equation}
where $u_{\rm rad}$ is the lab-frame energy density of the background
radiation and $\sigma_T$ is the Thomson cross section. To determine the
total inverse-Compton power per unit volume in the lab frame, we must
integrate Eq.~(A1) over the electron distribution as seen in that
frame. For simplicity, we assume that the electron distribution is isotropic
and monoenergetic as viewed in the frame of the jet. Naturally, in the lab
frame the electron distribution is neither monoenergetic nor isotropic due
to the Lorentz boost. Hence we must perform a Lorentz transformation of the
electron distribution in order to integrate Eq.~(A1). It is convenient
for this purpose to introduce the `electron intensity' $I_\epsilon$, which
is defined so that in time $dt$, the energy passing through a differential
cylinder with face area $dA$ due to electrons with energy in the range
$d\epsilon$ propagating in the differential solid angle $d\Omega$ is given by
$dE = I_\epsilon \, d\Omega \, dA \, dt \, d\epsilon$.
The total electron number density $n$ and energy density $u$ in the
lab frame are expressed
in terms of $I_\epsilon$ by the integrals
\begin{equation}
n = \int \int_0^\infty \frac{I_\epsilon}{c \epsilon} \,
d\epsilon \, d\Omega \ ,
\ \ \ \ \ \ 
u = \int \int_0^\infty \frac{I_\epsilon}{c} \,
d\epsilon \, d\Omega \ ,
%\eqno(\rm A2)
\end{equation}
respectively.

The electron Lorentz factor in the jet frame ($\gapprox 10^4$)
greatly exceeds the jet bulk Lorentz factor $\Gamma_j \lapprox 10$, and
therefore the electrons are ultrarelativistic in both the jet and lab frames.
It follows that the electron intensity transforms in exactly the same manner
as the usual radiation intensity. By analogy with the radiation case, we
can consequently relate the intensities in the two frames using
\begin{equation}
\frac{I_\epsilon(\epsilon)}{\epsilon^3}
= \frac{I_\epsilon'(\epsilon')}{\epsilon'^3} \ ,
%\eqno(\rm A3)
\end{equation}
where $\epsilon$ is the electron energy and primed quantities are
measured in the jet frame. The electron energies are likewise
related by the formula
\begin{equation}
\epsilon' = \epsilon \, \Gamma_j \, (1-\beta_j\cos\phi) \ ,
%\eqno(\rm A4)
\end{equation}
where $\phi$ is the angle of propagation of the electron relative to
the axis of the jet, as measured in the lab frame.

Since the electron distribution is isotropic and monoenergetic in the
jet frame, the electron intensity in that frame is given simply by
\begin{equation}
I_\epsilon'(\epsilon') = \frac{c \, \epsilon' \, n' \, \delta(\epsilon'-\epsilon_*)}
{4 \pi} \ ,
%\eqno(\rm A5)
\end{equation}
where
 $\epsilon_*$ is the energy of each electron in the jet frame and
$n'$ is the jet-frame electron number density. Utilizing Eqs.~(A3)--(A5), 
we can now express the lab-frame electron intensity as
\begin{equation}
I_\epsilon(\epsilon) = \frac{n' c \epsilon}{4 \pi \Gamma_j^2
(1-\beta_j\cos\phi)^2} \ \delta\left[\epsilon \Gamma_j(1-\beta_j\cos\phi)
-\epsilon_* \right] .
%\eqno(\rm A6)
\end{equation}
Before utilizing this result to calculate the lab-frame number and energy
densities, it is convenient to transform the $\delta$-function using
\begin{displaymath}
\delta\left[\epsilon \Gamma_j(1-\beta_j\cos\phi)-\epsilon_* \right] 
\end{displaymath}
\begin{equation}
\ \ \ \ \ = \delta\left[\epsilon - \frac{\epsilon_*}{ \Gamma_j(1-\beta_j\cos\phi)}
\right] \frac{1}{\Gamma_j(1-\beta_j\cos\phi)} \ .
%\eqno(\rm A7)
\end{equation}
Substituting this expression into Eqs.~(A2) and integrating over
$\epsilon$ yields
\begin{displaymath}
n =  \int \frac{n'}{4 \pi \Gamma_j^3 (1 - \beta_j \cos\phi)^3} \ d\Omega \ ,
\end{displaymath}
\begin{displaymath}
u = \int \frac{n' \epsilon_*}{4 \pi \Gamma_j^4 (1 - \beta_j \cos\phi)^4}
\ d\Omega \ .
\end{displaymath}
Setting $d\Omega = 2\pi \sin\phi \, d\phi$ and integrating over $\phi$
in the range $0 \le \phi \le \pi$, we obtain for the lab-frame number
and energy densities the exact results
\begin{equation}
n =  n' \, \Gamma_j \ ,
\ \ \ \ \ \ \ 
u = u' \, \Gamma_j^2 \left(1 + \frac{1}{3}\beta_j^2 \right) \ ,
%\eqno(\rm A8)
\end{equation}
where $u' = n' \epsilon_*$ is the jet-frame electron energy density.

Next we calculate the total power per unit volume emitted by the electrons
due to inverse-Compton scattering of the background radiation, given by
\begin{equation}
\dot u = \int \int_0^\infty P_{\rm compt}(\gamma) \, \frac{I_\epsilon}{c \epsilon}
\, d\epsilon \, d\Omega \ ,
%\eqno(\rm A9)
\end{equation}
where $\gamma=\epsilon/(m_e c^2)$ and $m_e$ is the electron rest mass.
Substitution using Eqs.~(A1), (A6), and (A7) yields, upon integration
\begin{equation}
\dot u = \frac{4}{3} \, \frac{c \, \sigma_T \, u_{\rm rad}}{ m_e^2 c^4}
\ u' \, \epsilon_* \,
\Gamma_j^3 \, (1 + \beta_j^2) \ .
%\eqno(\rm A10)
\end{equation}
Note that in deriving this expression we have set $\beta=1$ in Eq.~(A1),
which is an extremely accurate approximation for the ultrarelativistic electrons
of interest here. Our final step is to compute the inverse-Compton loss
timescale, $\tau \equiv u / \dot u$, given by
\begin{equation}
\tau =  \frac{3 \, m_e \, c^2}{4 \, c \, \sigma_T \, u_{\rm rad}} \,
\frac{1}{\gamma_* \, \Gamma_j} \, \left[\frac{1 + \beta_j^2/3}
 {1 + \beta_j^2}\right] \ ,
%\eqno(\rm A11)
\end{equation}
where $\gamma_* \equiv \epsilon_*/(m_e c^2)$ is the Lorentz factor of the
monoenergetic electrons in the jet frame (denoted
$\gamma$ in the main text). Eq.~(A11) gives the value
of $\tau$ measured in the lab frame by virtue of the definitions of $u$
and $\dot u$, and the factor in brackets 
provides the correction factor quoted following Eq.~(6).}


\begin{thebibliography}{}


\bibitem[2000] {arsh} Arshakian, T.\ G., \& Longair, M.\ S. 2000, MNRAS, 311, 846
\bibitem[1989] {bart} Barthel, P.\ D. 1989, ApJ, 336, 606
\bibitem[1984] {bege84} Begelman, M.\ C., Blandford, R.\ D., \& Rees, M.\ J. 1984, 
Rev.\ Mod.\ Phys., 56, 255
\bibitem[1994] {bege94} Begelman, M.\ C., Rees, M.\ J., \& Sikora, M. 1994, ApJ, 429, L57
\bibitem[1978] {bell} Bell, A.\ R. 1978, MNRAS, 182, 147
\bibitem[1995] {bick} Bicknell, G.\ V. 1995, ApJS, 101, 29
\bibitem[1987] {bier} Biermann, P.\ L., \& Strittmatter, P.\ A. 1987, ApJ, 322, 643
\bibitem[1999] {bire} Biretta, J.\ A., Sparks, W.\ B., \&
 Macchetto, F. 1999, ApJ,  520, 621 
\bibitem[1974] {blan} Blandford, R.\ D., \& Rees, M.\ J. 1974, MNRAS, 169, 395
\bibitem[1996] {brid96} Bridle, A.\ H., 1996, in Energy Transport in Extragalactic
 Radio Sources, ed.\ P.\ E.\  Hardee, A.\ H.\ Bridle, \& J.\ A.\ Zensus
(ASP, San Francisco), 383
\bibitem[1994] {brid94} Bridle, A.\ H., Hough, D.\ H., Lonsdale, C.\ J., 
Burns, J.\ O., \& Laing, R.\ A. 1994, AJ, 108, 766
\bibitem[1996] {cari} Carilli, C.\ L., Perley, R.\ A., Bartel, N., \& 
Sorathia, B. 
1996,  in Energy Transport in Extragalactic
 Radio Sources, ed.\ P.\ E.\  Hardee, A.\ H.\ Bridle, \& J.\ A.\ Zensus
(ASP, San Francisco), 287
\bibitem[2000a] {chiaa} Chiaberge, M., Capetti, A., \& Celotti, A. 2000a, A\&A, 355, 873
\bibitem[2000b] {chiab} Chiaberge, M., Celotti, A., Capetti, A., \& Ghisellini,
G. 2000b, A\&A, 358, 104
\bibitem[1995] {cont} Contopoulos, J., \& Kazanas, D. 1995, ApJ, 441, 521
\bibitem[1983] {cran} Crane, P., Tyson, J.\ A., \& Saslaw, W.\ C. 1983, ApJ, 265, 681
\bibitem[1987] {dar} Dar, A., \& Laor, A. 1987, ApJ, 478, L5
\bibitem[1993] {derm93} Dermer, C.\ D.,  \& Schlickeiser, R. 1993, ApJ, 416, 458
\bibitem[1994] {derm94} Dermer, C.\ D.,  \& Schlickeiser, R., 1994, ApJS, 90, 945
\bibitem[1986] {dreh} Dreher, J.\ W., Simkin, S.\ M. 1986, AJ, 91, 58
\bibitem[1983] {drur} Drury, L.O'C., 1983, Rep Prog Phys 46, 973
\bibitem[1978] {eich} Eichler, D., \& Wiita, P.\ J. 1978, Nat, 274, 38
\bibitem[1997] {eile} Eilek, J.\ A., Melrose, D.\ B., \& Walker, M.A. 1997, ApJ, 483, 282
\bibitem[1974] {fana} Fanaroff, B.\ L., \& Riley, J.\ M. 1974, MNRAS, 167, 31P 
\bibitem[1968] {felt} Felton, J.\ E. 1968, ApJ, 151, 861
\bibitem[1988] {garr} Garrington, S.\ T., Leahy, J.\ P., Conway, R.\ G.,   
Laing, R.\ A.,
1988, Nat, 331, 147
\bibitem[2001] {ghis} Ghisellini, G., \& Celloti, A. 2001, in Issues of Unifications of AGNs,
ed.\ R.\ Maiolino, A.\ Marconi, \& N.\ Nagar, in press (astro-ph/0108110)
\bibitem[1999]{giov} Giovannini, G., Taylor, G.\ B.,
 Arbizzani, E., et al. 1999, ApJ, 522, 101
\bibitem[1980] {gk} Gopal-Krishna 1980, A\&A, 81, 328
\bibitem[1996] {gkkw} Gopal-Krishna, Kulkarni, V.\ K., \& Wiita, P.\ J. 1996, ApJ, 463, L1
\bibitem[2001] {hard} Hardcastle, M.\ J., Birkinshaw, M., \& Worrall, D.\ M.
2001, MNRAS, in press, astro-ph/0101240
\bibitem[1976] {harg} Hargrave, R.\ J., \& Ryle, M. 1976, MNRAS, 175, 481
\bibitem[2001] {harr01} Harris, D.\ E. 2001, in  Particles
and Fields in Radio Galaxies, ed.\ R.\ A.\ Laing \& K.\ A.\
 Blundell,   (ASP, San Francisco), in press
\bibitem[1994] {harr94} Harris, D.\ E., Carilli, C.\ L., \& Perley, R.\ A. 1994, 
Nature, 367, 713
\bibitem[1998] {harr98} Harris, D.\ E., Leighly, K.\ M., \& Leahy, J.\ P. 1998, ApJ, 499, L149
\bibitem[2000] {harr00} Harris, D.\ E., Nulsen, P.\ E.\ J., Ponman, T.\ J., et al.
 2000, ApJ, 530, L81
\bibitem[1987] {heav} Heavens, A.\ F., \& Meisenheimer, K. 1987, MNRAS, 225, 335
\bibitem[1997] {hein} Heinz, S., \&  Begelman, M.\ C. 1997, ApJ, 490, 653
\bibitem[1994] {hilt} Hiltner, P.\ R., Meisenheimer, K., R\"oser, H.-J., 
Laing, R.\ A., \&
Perley, R.A. 1994, A\&A, 286, 25
\bibitem[1977] {jenk} Jenkins, C.\ J., Pooley, G.\ G., \& Riley, J.\ M.
 1977, Mem.\ RAS, 84, 61
\bibitem[1988] {keel88} Keel, W.\ C. 1988, ApJ, 329, 532
\bibitem[1995] {keel95} Keel, W.\ C., \& Martini, P. 1995, AJ, 109, 2305
\bibitem[1987] {kirk} Kirk, J.\ G., \& Schneider, P. 1987, ApJ, 322, 256
\bibitem[1990] {komi} Komissarov, S.\ S. 1990, Sov.\ Astr.\ Lett., 16, 284
\bibitem[1976] {kron76} Kronberg, P. 1976, ApJ, 203, L47
\bibitem[1977] {kron77} Kronberg, P., van den Bergh, S., \& Button, S. 1977, AJ, 82, 1039
\bibitem[1986] {kund86} Kundt, W. 1986, in Astrophysical Jets and Their
Engines, ed.\ W.\ Kundt  (Kluwer, Dordrecht),  1
\bibitem[1996] {kund96} Kundt, W. 1996, in Jets from Stars 
and Galactic Nuclei, ed.\ W.\ Kundt, (Springer, Berlin),  1
\bibitem[1999] {laht} L\"ahteenm\"aki. A., \& Valtaoja, E. 1999, AJ, 117, 1168
\bibitem[1988] {lain88} Laing, R.\ A. 1988, Nature, 331, 149
\bibitem[1993] {lain93} Laing, R.\ A., 1993, in STScI Symp. 6, Astrophysical
Jets, ed.\ D.\ Burgarella, M.\ Livio, C.\ O'Dea, 
(Cambridge University Press), 95
\bibitem[1999] {lain99} Laing, R.\ A., Parma, P., de Ruiter, H.\ R., \&  Fanti, R.
   1999, MNRAS, 306, 513
\bibitem[1999] {lang} Lang, K.\ R. 1999, Astrophysical Formulae (Springer, New York)
\bibitem[1995] {leah} Leahy, J.\ P., \& Perley, R.\ A. 1995, MNRAS, 277, 1097
\bibitem[1975] {leli} Lelievre, G., \& Wlerick, G. 1975, A\&A, 42, 293
\bibitem[1984] {linf} Linfield, R., \& Perley, R. 1984, ApJ, 279, 60
\bibitem[1981] {long} Longair, M.\ S., 1981, High Energy Astrophysics, (Cambridge
University Press), 279
\bibitem[2000] {loon} Looney, L.\ W., \& Hardcastle, M.\ J. 2000, ApJ, 534, 172
\bibitem[1993] {mann93} Mannheim, K. 1993, A\&A, 269, 67
\bibitem[1991] {mann91} Mannheim, K., \& Biermann, P.\ L., \& Kr\"ulls, W.\ M. 
1991, A\&A, 251, 723
\bibitem[1995] {mast} Mastichiadis, A., \& Kirk, J.\ G. 1995, A\&A, 295, 613
\bibitem[1996] {meis96} Meisenheimer. K. 1996, in Jets from Stars 
and Galactic Nuclei, ed.\ W.\ Kundt, (Springer, Berlin), 57
\bibitem[1986] {meis86} Meisenheimer, K., \& R\"oser, H.-J. 1986, Nature, 319, 459
\bibitem[1989] {meis89} Meisenheimer, K., R\"oser, H.-J., Hiltner, P.\ R., 
Yates, M.\ G., Longair, M.\ S., Chini, R., \& Perley, R.\ A. 
1989, A\&A, 219, 63
\bibitem[1997] {meis97} Meisenheimer, K., Yates, M.\ G., \& R\"oser, H.-J.
 1997, A\&A, 325, 57
\bibitem[1992] {okay} Okayasu, R., Ishiguro, M., \& Tabara, H. 1992, PASJ, 44, 335
\bibitem[1982] {orr} Orr, M.\ J.\ L.,  \& Browne, I.\ W.\ A. 1982, MNRAS, 200, 1067
\bibitem[1989] {owen} Owen, F.\ N., Hardee, P.\ E.,  \& Cornwell, T.\ J. 
1989, ApJ, 340, 698 
\bibitem[1984]{perl84} Perley, R.\ A., Dreher, D.\ W., \& Cowan, J.\ J. 1984, ApJ, 285, L35
\bibitem[1997] {perl97} Perley, R.\ A., R\"oser, H.-J., \& Meisenheimer, K. 
1997, A\&A, 328, 12
\bibitem[1995] {pola} Polatidis, A.\ G., Wilkinson, P.\ N., Xu, W., 
Readhead, A.\ C.\ S.,
Pearson, T.\ J., Taylor, G.\ B., \& Vermeulen, R.\ C. 1995, ApJS, 98, 1
\bibitem[1997] {prie} Prieto, M.\ A.,  \& Kotilainen, J.\ K. 1997, ApJ, 491, L77
\bibitem[1967] {rees67} Rees, M.\ J. 1967, MNRAS, 135, 345
\bibitem[1971] {rees71} Rees, M.\ J. 1971, Nature, 229, 312
\bibitem[1997] {ridg} Ridgway, S.\ E., \& Stockton, A. 1997, AJ, 114, 511
\bibitem[1987] {rose87} R\"oser, H.-J., \& Meisenheimer, K. 1987, ApJ, 314, 70
\bibitem[1991] {rose91} R\"oser, H.-J., \& Meisenheimer, K. 1991, A\&A, 252, 458
\bibitem[2000] {rose00} R\"oser, H.-J., Meisenheimer, K., Neumann, M., Conway,
R.\ G., \& Perley, R.\ A. 2000, A\&A, 360, 99
\bibitem[1979] {rybi} Rybicki, G.\ B., \& Lightman, A.\ P. 1979, Radiative Processes
in Astrophysics (Wiley, New York)
\bibitem[1987] {saik} Saikia, D.\ J., Salter, C.\ J., \& Muxlow, T.\ W.\ B.
 1987, MNRAS, 224, 911
\bibitem[1978] {sasl} Saslaw, W.\ C., Tyson, A., \& Crane, P. 1978, ApJ, 222, 435
\bibitem[1987] {sche87} Scheuer, P.\ A.\ G. 1987, in Astrophysical Jets and their
Engines, ed.\ W.\ Kundt, (Reidel, Dordrecht), 129
\bibitem[1989] {sche89} Scheuer, P.\ A.\ G. 1989, in Hot Spots in Extragalactic  
Radio Sources, ed.\ K.\ Meisenheimer \& H.-J.\ R\"oser, 
(Springer, Berlin),  159 
\bibitem[1995] {sche95} Scheuer, P.\ A.\ G. 1995, MNRAS, 277, 331
\bibitem[2000] {scot} Scott, D., Silk, J., Kolb, E.\ W., \& Turner, M.\ S. 
2000, in 
Allen's Astrophysical Quantities, ed.\ A.\ N.\ Cox (Springer, New
York),  664
\bibitem[1984] {shkl} Shklovskii, I.\ S. 1984, Sov.\ Astr., 28, 489 
\bibitem[1990] {siah} Siah, M.\ J., \& Wiita, P.\ J. 1990, ApJ, 363, 411
\bibitem[1978] {simk78} Simkin, S. 1978, ApJ, 222, L55 
\bibitem[1979] {simk79} Simkin, S. 1979, ApJ, 234, 56 
\bibitem[1999] {subr} Subramanian, P., Becker, P.\ A., \&  Kazanas, D. 1999, ApJ, 523, 203 
\bibitem[1998] {swai} Swain, M.\ R., Bridle, A.\ H., \& Baum, S.\ A. 1998, ApJ, 507, L29
\bibitem[1995] {thom} Thomson, R.\ C., Crane, P., \&  Mackay, C.\ D. 1995, ApJ, 446, L93
\bibitem[1993] {trib} Tribble, P.\ C. 1993, MNRAS, 261, 57
\bibitem[2000] {valt} Valtonen, M.\ J., \& Hei{}n\"am\"aki, P. 2000, ApJ, 530, 107
\bibitem[1994] {verm} Vermeulen, R.\ C., \& Cohen, M.\ H. 1994, ApJ, 430, 467
\bibitem[1997] {ward} Wardle, J.\ F.\ C., \& Aaron, S.\ E.  1997, MNRAS, 286, 425
\bibitem[2000] {wils00} Wilson, A.\ S., Young, A.\ J., \& Shopbell, P.\ L. 2000, ApJ, 544, L27
\bibitem[2001] {wils01} Wilson, A.\ S., Young, A.\ J., \& Shopbell, P.\ L. 2001, ApJ, 547, 740


\end{thebibliography}
\end{document}